\documentclass[pra,twocolumn,showpacs]{revtex4}
\usepackage{epsfig}
\usepackage{graphicx}
\usepackage{dcolumn}
\usepackage{amsmath,amssymb}
\usepackage{pstricks}
\usepackage{color}
\usepackage{footnote}
\usepackage{bm}
\usepackage{bbold}
\usepackage[colorlinks=true,citecolor={blue},urlcolor={blue}, linkcolor=blue]{hyperref}
\begin{document}
\title{Relativistic coupled-cluster study of RaF as a candidate for parity and time reversal violating interaction}
\author{Sudip Sasmal,$^{1,}$\footnote{sudipsasmal.chem@gmail.com}
Himadri Pathak,$^{1,}$\footnote{hmdrpthk@gmail.com}
Malaya K. Nayak,$^{2,}$\footnote{mk.nayak72@gmail.com}
Nayana Vaval,$^1$ Sourav Pal$^{3}$}
\affiliation{$^1$Electronic Structure Theory Group, Physical Chemistry Division, CSIR-National Chemical Laboratory, Pune, 411008, India}
\affiliation{$^2$Theoretical Chemistry Section, Bhabha Atomic Research Centre, Trombay, Mumbai 400085, India}
\affiliation{$^3$Department of Chemistry, Indian Institute of Technology Bombay, Powai, Mumbai 400076,  India}
\begin{abstract}
We have employed both Z-vector method and the expectation value approach in the relativistic coupled-cluster framework to calculate the scalar-pseudoscalar
(S-PS) ${\mathcal{P,T}}$-odd interaction constant ($W_\text{s}$) and the effective electric field ($E_\text{eff}$) experienced by the
unpaired electron in the ground electronic state of RaF.
Further, the magnetic hyperfine structure constants of $^{223}$Ra in RaF and $^{223}$Ra$^{+}$ are also calculated
and compared with the experimental values wherever available to judge
the extent of accuracy obtained in the employed methods.
The outcome of our study reveals that the Z-vector method is superior than the expectation value approach in terms of accuracy obtained 
for the calculation of ground state property. The Z-vector calculation shows that RaF has a high $E_\text{eff}$ (52.5 GV/cm)
and $W_\text{s}$ (141.2 kHz) which makes it a potential candidate for the eEDM experiment.
\end{abstract}
\maketitle
\section{Introduction}
The ongoing accelerator based experiments in the search for new physics can solve some of the unanswered problems of the fundamental physics
like matter-antimatter asymmetry. A complementary to these high energy experiments is the 
search for violation in spatial inversion (${\mathcal{P}}$) and time reversal (${\mathcal{T}}$) symmetries
in nuclei, atoms or molecules in the low energy domain using non-accelerator
experiments \cite{ginges_2004, sandars_1965, sandars_1967, labzovskii_1978, barkov_1980, shapiro_1968, pospelov_2005}.
One of such ${\mathcal{P,T}}$-violating interaction results
into the electric dipole moment of electron (eEDM) \cite{bernreuther_1991, tl_edm, ybf_edm, tho_edm}. The eEDM predicted by the standard model (SM) of
elementary particle physics is too small ($< 10^{-38}$ e cm) \cite{khriplovich_2011} to be observed by the today's experiment.
However, many extensions of the SM predict the value of eEDM to be in the range of $10^{-26} - 10^{-29}$ e cm \cite{commins_1999}
and the sensitivity of the modern eEDM experiment also lies in the same range.
Till date, the experiment done by ACME collaboration \cite{tho_edm} using ThO yields the best upper bound limit of eEDM.
The high sensitivity of modern
eEDM experiment is mainly due to the fact that heavy paramagnetic diatomic molecules
offer a very high internal effective electric field ($E_\mathrm{eff}$), which enhances the eEDM effects \cite{sushkov_1978, flambaum_1976}.
In the experiment, both eEDM and the coupling interaction between the scalar-hadronic current and the pseudoscalar
electronic current contribute to the P,T-odd frequency shift. Therefore, it is impossible to
decouple the individual contribution from these two effects in a single experiment. However, it is possible to untwine these two contributions from each other
and an independent limit on the value of eEDM ($d_e$) and scalar-pseudoscalar (S-PS) coupling constant ($k_s$) can be obtained
by using data from two different experiments as suggested by Dzuba {\it et al} \cite{dzuba_2011}.
It is, therefore, an accurate value of the $E_\mathrm{eff}$ and the scalar-pseudoscalar (S-PS) ${\mathcal{P,T}}$-odd interaction
constant ($W_\mathrm{s}$) are needed since these two quantities cannot be measured by means of any experiment.
Therefore, one has to rely on an accurate {\it ab initio} theory that can simultaneously take care of the effects of relativity
and electron correlation for the calculation of these quantities.\par
%
The best way to include the effects of special relativity in the electronic structure calculations is to solve the
Dirac-Hartree-Fock (DHF) equation in the four-component framework. The DHF method considers an average
electron-electron interaction and thus misses the correlation between electrons having same spin. On the other hand,
the single reference coupled-cluster (SRCC) method is the most preferred many-body theory to incorporate the dynamic
part of the electron correlation.
The calculation of property in the SRCC framework can be done either numerically or analytically. 
In numerical method (also known as the finite-field (FF) method), the coupled-cluster amplitudes are
functions of the external field parameters \cite{monkhorst_1977} and thus, for calculations of each property, separate set of CC calculation
is needed. The error associated with the FF method is also dependent on the method of calculation, i.e., the
number of data points considered for the numerical differentiation.
On the contrary, in the analytical method, the CC amplitudes are independent of the external field of perturbation and therefore,
one needs to solve only one set of CC equation for the calculations of any number of properties.
Normal CC (NCC) method being non-variational, does not satisfy the generalized Hellmann-Feynman (GHF) theorem and thus,
the expectation value and the energy derivative approach are two different formalisms for the calculation of first order property.
However, the energy derivative in NCC framework is the corresponding expectation value plus some additional terms which
make it closer to the property value obtained in the full configuration interaction (FCI) method. Thus, the property value obtained in the energy
derivative method is much more reliable than the corresponding expectation value method. Another disadvantage of the expectation
value method is that it leads to a non-terminating series and any truncation further introduces an additional error.
The Z-vector method \cite{schafer_1984, zvector_1989} (an energy derivative method), on the other hand, leads to a naturally terminating series
at any level of approximation.
The higher order derivative in the NCC framework can be calculated by using the Lagrange multiplier method \cite{koch_1990}
and for the first order energy derivative, it leads to the identical equations as of Z-vector
method.
It is worth to note that there are alternative options like
expectation value CC (XCC) \cite{bartlett_xcc}, unitary CC (UCC) \cite{bartlett_ucc}, and
extended CC (ECC) \cite{arponen_ecc, bishop_ecc} to solve the SRCC equation. All these methods are known in the literature as
the variational coupled-cluster (VCC) method \cite{szalay_1995}.
These VCC methods are well established in the non-relativistic framework but are not that much popular in the relativistic domain, a few are documented
in the literature like relativistic UCC by Sur {\it et al.} \cite{mukherjee_ucc,rajat_ucc}, applicable only for the purpose of atomic calculations.
Recently, Sasmal {\it et al} implemented ECC   
in the four-component relativistic domain
to calculate the magnetic hyperfine structure (HFS) constants of both atoms and molecules in their open-shell ground state configuration \cite{sasmal_ecc}.
The ECC method being variational satisfies the GHF theorem, therefore, expectation value and the energy derivative approach are identical to each other.
However, in ECC method amplitude equations for the excitation and de-excitation
operators are coupled to each other, whereas, in Z-vector method, the amplitude equations of excitation operator are decoupled
from the amplitude equations of the de-excitation operator. This accelerates the convergence in the Z-vector method
with a lesser computational cost as compared to the ECC. \par
In this work, we have calculated the $E_\text{eff}$ and $W_\text{s}$ of RaF in its ground electronic
($^2\Sigma$) state using Z-vector method in the CC framework. We also calculated these properties in the expectation
value method to show the superiority of the Z-vector method over the expectation value method.
We have chosen the RaF molecule for the following reasons: This molecule has been proposed for the
${\mathcal{P}}$-odd and ${\mathcal{P,T}}$-odd experiment \cite{isaev_2010, isaev_2013, kudashov_2014} due to its high Schiff moment, $E_\mathrm{eff}$
and $W_\text{s}$.
The $E_\text{eff}$ of $^2\Sigma$ state of
RaF is even higher than the ground state ($^2\Sigma$) of YbF. Therefore, the more precise value of $E_\text{eff}$
and $W_\text{s}$ and their ratio are very important for the eEDM experiment using this molecule. RaF
can be directly laser cooled as it has high diagonal Franck-Condon matrix element between the ground
and first excited electronic state and the corresponding transition frequency lies in the visible
region with a reasonable lifetime \cite{isaev_2010}. \par
The manuscript is organized as follows. A brief overview of the expectation value and the Z-vector method
in the CC framework including concise details of the properties calculated in this work are given in Sec. \ref{theory}.
Computational details are given in Sec. \ref{comp}. We presented our calculated results and
discuss about those in Sec. \ref{res_dis} before making concluding remark.
Atomic unit is used consistently unless stated.
\section{Theory}\label{theory}
\subsection{Expectation value and Z-vector method}\label{corr}
The DHF wavefunction is the best description of the ground state in a single determinant theory and thus, it is
used as a reference function for the correlation calculations where the Dirac-Coulomb (DC)
Hamiltonian is used which is given by
\begin{eqnarray}
{H_{DC}} &=&\sum_{i} \Big [-c (\vec {\alpha}\cdot \vec {\nabla})_i + (\beta -{\mathbb{1}_4}) c^{2} + V^{nuc}(r_i)+ \nonumber\\
       && \sum_{j>i} \frac{1}{r_{ij}} {\mathbb{1}_4}\Big].
\end{eqnarray}
Here, {\bf$\alpha$} and $\beta$ are the usual Dirac matrices, $c$ is the speed of light,
${\mathbb{1}_4}$ is the 4$\times$4 identity matrix and the sum is over all the electrons, which is denoted by
$i$. The Gaussian charge distribution is used as nuclear potential function ($V^{nuc}(r_i)$).
The DHF method approximates the electron-electron repulsion in an average way and thus misses
the correlation between same spin electrons. In this article, we have used the SRCC method to incorporate
the dynamic part of electron correlation. The SRCC wavefunction is given by
$|\Psi_{cc}\rangle=e^{T}|\Phi_0\rangle$ ,
where $\Phi_0$ is the DHF wavefunction and $T$ is coupled-cluster excitation operator which is given
by
\begin{eqnarray}
 T=T_1+T_2+\dots +T_N=\sum_n^N T_n ,
\end{eqnarray}
with
\begin{eqnarray}
 T_m= \frac{1}{(m!)^2} \sum_{ij\dots ab \dots} t_{ij \dots}^{ab \dots}{a_a^{\dagger}a_b^{\dagger} \dots a_j a_i} ,
\end{eqnarray}
where i,j(a,b) are the hole(particle) indices and $t_{ij..}^{ab..}$ are the cluster amplitudes corresponding 
to the cluster operator $T_m$.
In coupled-cluster single and double (CCSD) model, $T=T_1+T_2$. The equations for T$_1$ and T$_2$ are
given as
\begin{eqnarray}
 \langle \Phi_{i}^{a} | (H_Ne^T)_c | \Phi_0 \rangle = 0 , \,\,
  \langle \Phi_{ij}^{ab} | (H_Ne^T)_c | \Phi_0 \rangle = 0 ,
 \label{cc_amplitudes}
\end{eqnarray}
where H$_N$ is the normal ordered DC Hamiltonian and subscript $c$ means only the connected terms exist in the
contraction between H$_N$ and T. Size-extensivity is ensured by this connectedness. \par
\begin{figure}[ht]
\centering
\begin{center}
\includegraphics[scale=.1, height=4.0cm]{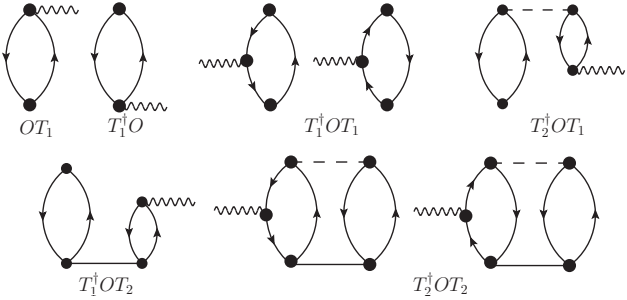}
\caption {Diagrams for expectation value approach using linear truncation scheme}
\label{lin_expec}
\end{center}  
\end{figure}

Once the cluster amplitudes are solved, the expectation value of any property operator of interest, $\langle O_N \rangle$ can be calculated by the following expression as given in Ref. \cite{cizek_1967},
\begin{eqnarray}
\langle O_N \rangle=\frac{\langle \Psi_{cc} | O_N | \Psi_{cc} \rangle}{\langle \Psi_{cc} | \Psi_{cc} \rangle}
 &=& \frac{\langle \Phi_0 e^{T^{\dagger}} | O_N | e^{T} \Phi_0 \rangle}{\langle \Phi_0 | e^{T^{\dagger}} e^{T} | \Phi_0 \rangle} \nonumber\\
 &=& \langle \Phi_0 | (e^{T^{\dagger}} O_N e^{T})_c | \Phi_0 \rangle.
\end{eqnarray}
The above series is a non-terminating series. Since, the dominant contribution comes from
the linear terms, therefore, linear approximation is the most favored choice.
The detailed diagrammatic expression considering only linear terms within the CCSD approximation is given in
Fig. \ref{lin_expec} and the corresponding algebraic equation is given as in Eq. \ref{expect_eqn}.
We have used Einstein summation convention, i.e., the repeated indices are summed over in the expression.
The $t$ amplitudes with particle(hole) indices at the subscript(superscript) are the corresponding amplitudes
of the $T^{\dagger}$ operator. It is interesting to note that there is no possible diagrams (as well as algebraic
expression) of the kind $T_2^{\dagger}O$ or $OT_2$, since closed connected diagrams can not be constructed by 
these two expressions.
%
\begin{widetext}
\begin{eqnarray}
\langle O \rangle &=& O(i,a) \cdot t_{i}^{a} + t_{a}^{i} \cdot O(a,i) + t_{a}^{i} \cdot O(a,b) \cdot t_{i}^{b} 
 - t_{a}^{i} \cdot O(j,i) \cdot t_{j}^{a} + t_{ab}^{ij} \cdot O(b,j) \cdot t_{i}^{a}+\nonumber \\
    && t_{a}^{i} \cdot O(j,b) \cdot t_{ij}^{ab}-\frac{1}{2} t_{ab}^{ij} \cdot O(k,j) \cdot t_{ik}^{ab} + \frac{1}{2} t_{ab}^{ij} \cdot O(b,c) \cdot t_{ij}^{ac} .
\label{expect_eqn}
\end{eqnarray}
\end{widetext}

\par
The CC amplitudes are solved in a nonvariational way (using Eq. \ref{cc_amplitudes}) and thus, the CC energy
is not minimized with respect to the determinantal coefficient and the molecular orbital coefficient 
in the expansion of the many electron correlated wavefunction
for a fixed nuclear geometry \cite{monkhorst_1977}. Therefore, the calculation of CC energy
derivative needs to include the derivative of energy with respect to these two coefficients in addition to the derivative of
these two parameters with respect to the external field of perturbation.
\begin{figure}[ht]
\centering
\begin{center}
\includegraphics[scale=.1, height=6.0cm]{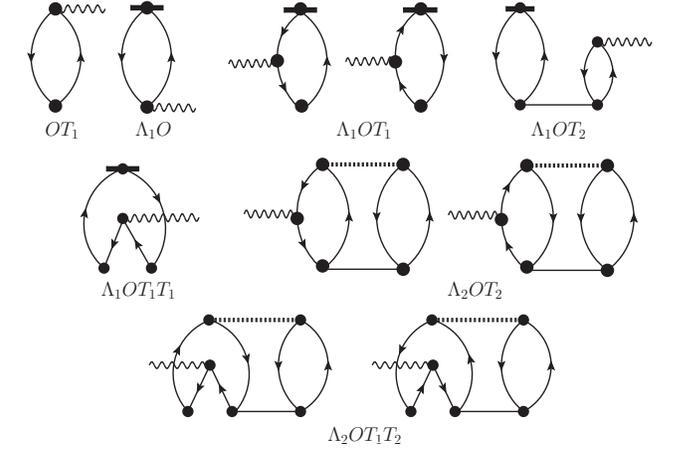}
\caption {Diagrams for the energy derivative in Z-vector method}
\label{z_vec_prop}
\end{center}  
\end{figure}
However, the derivative terms associated with determinantal coefficient can be integrated by
the introduction of a perturbation independent linear operator, $\Lambda$ \cite{zvector_1989}. $\Lambda$ is an antisymmetrized
de-excitation operator whose second quantized form is given by 
\begin{eqnarray}
 \Lambda=\Lambda_1+\Lambda_2+ \dots+\Lambda_N=\sum_n^N \Lambda_n ,
\end{eqnarray}
where
\begin{eqnarray}
 \Lambda_m= \frac{1}{(m!)^2} \sum_{ij \dots ab \dots} \lambda_{ab \dots}^{ij \dots}{a_i^{\dagger}a_j^{\dagger} \dots a_b a_a} ,
\end{eqnarray}
where $\lambda_{ab \dots}^{ij \dots}$ are the cluster amplitudes corresponding  to the operator $\Lambda_m$.
The detailed description of $\Lambda$ operator and amplitude equation is given in Ref. \cite{zvector_1989}.
In CCSD model, $\Lambda=\Lambda_1+\Lambda_2$. The explicit equations for the amplitudes of $\Lambda_1$
and $\Lambda_2$ operators are given by
\begin{eqnarray}
\langle \Phi_0 |[\Lambda (H_Ne^T)_c]_c | \Phi_{i}^{a} \rangle + \langle \Phi_0 | (H_Ne^T)_c | \Phi_{i}^{a} \rangle = 0,
\end{eqnarray}
\begin{eqnarray}
\langle \Phi_0 |[\Lambda (H_Ne^T)_c]_c | \Phi_{ij}^{ab} \rangle + \langle \Phi_0 | (H_Ne^T)_c | \Phi_{ij}^{ab} \rangle \nonumber \\
 + \langle \Phi_0 | (H_Ne^T)_c | \Phi_{i}^{a} \rangle \langle \Phi_{i}^{a} | \Lambda | \Phi_{ij}^{ab} \rangle = 0.
\label{lambda_2}
\end{eqnarray}
In is interesting to note that the third term of Eq. \ref{lambda_2} is of the nature of disconnected type and
it eventually produces one disconnected diagram in the $\Lambda_2$ amplitude equation (for details see Ref. \cite{zvector_1989, sasmal_pra_rapid}).
Although the diagram is disconnected but it does not have any closed part. This ensures that the corresponding energy
diagram is linked which restores the size extensivity.
The energy derivative can be given as
\begin{eqnarray}
 \Delta E' = \langle \Phi_0 | (O_Ne^T)_c | \Phi_0 \rangle + \langle \Phi_0 | [\Lambda (O_Ne^T)_c]_c | \Phi_0 \rangle
\end{eqnarray}
where, $O_N$ is the derivative of normal ordered perturbed Hamiltonian with respect to external field of perturbation.
The detailed diagrammatic expression is given in Fig. \ref{z_vec_prop} and the corresponding algebraic equation is given in the following Eq. \ref{z_vec_prop_eqn},
\begin{widetext}
\begin{eqnarray}
 \Delta E'&=& O(i,a) \cdot t_{i}^{a} + \lambda_{a}^{i} \cdot O(a,i) + \lambda_{a}^{i} \cdot O(a,b) \cdot t_{i}^{b} 
  + \lambda_{a}^{i} \cdot O(j,i) \cdot t_{j}^{a} + \lambda_{a}^{i} \cdot O(j,b) \cdot t_{ij}^{ab}
 - \lambda_{a}^{i} \cdot O(j,b) \cdot t_{i}^{b} \cdot t_{j}^{a}- \nonumber \\
  && \frac{1}{2} \lambda_{ab}^{ij} \cdot O(k,j) \cdot t_{ik}^{ab}
+ \frac{1}{2} \lambda_{ab}^{ij} \cdot O(b,c) \cdot t_{ij}^{ac} - \frac{1}{2} \lambda_{bc}^{ik} \cdot O(j,a) \cdot t_{i}^{a} \cdot t_{jk}^{bc} 
- \frac{1}{2} \lambda_{ac}^{jk} \cdot O(i,b) \cdot t_{i}^{a} \cdot t_{jk}^{bc}.
\label{z_vec_prop_eqn}
\end{eqnarray}
\end{widetext}
\subsection{One electron property operators}\label{prop}
The $E_{\text{eff}}$ can be obtained by evaluating the following matrix element
\begin{eqnarray}
 E_{\text{eff}} = |W_d \Omega|  = | \langle \Psi_{\Omega} | \sum_j^n \frac{H_d(j)}{d_e} | \Psi_{\Omega} \rangle |,
 \label{E_eff}
\end{eqnarray}
where $\Omega$ is the component of total angular momentum along the molecular axis and $\Psi_{\Omega}$ is the
wavefunction corresponding to $\Omega$ state. $n$ is the total number of electrons and 
H$_d$ is the interaction Hamiltonian of $d_e$ with internal electric field and is given by \cite{kozlov_1987, titov_2006},
\begin{eqnarray}
 H_d = 2icd_e \gamma^0 \gamma^5 {\bf \it p}^2 ,
\label{H_d}
\end{eqnarray}
where $\gamma$ are the usual Dirac matrices and {\bf \it p} is the momentum operator.
\begin{table*}[ht]
\caption{ Cutoff used and correlation energy of the ground state of Ra$^{+}$ and RaF in different basis sets }
\begin{ruledtabular}
\newcommand{\mc}[3]{\multicolumn{#1}{#2}{#3}}
\begin{center}
\begin{tabular}{lccccccccr}
\mc{4}{c}{Basis} & \mc{2}{c}{Cutoff (a.u.)} & \mc{2}{c}{Spinor} & \mc{2}{c}{Correlation Energy (a.u.)}\\
\cline{1-4} \cline{5-6} \cline{7-8} \cline{9-10}
Name & Nature & Ra & F & Occupied & Virtual & Occupied & Virtual & MBPT(2) & CCSD \\
\hline
Ra$^{+}$ \\
A & TZ & dyall.cv3z & × & -30  & 500 & 51 & 323 & -1.74841495 & -1.57235409 \\
B & TZ & dyall.cv3z & × & -130 & 500 & 69 & 323 & -2.42790147 & -2.20700361 \\
C & TZ & dyall.cv3z & × &      & 500 & 87 & 323 & -2.78897499 & -2.55468917 \\
D & QZ & dyall.cv4z & × & -30  & 20  & 51 & 349 & -1.43221422 & -1.31515023 \\
E & QZ & dyall.cv4z & × & -130 & 20  & 69 & 349 & -1.49747209 & -1.37242346 \\
F & QZ & dyall.cv4z & × &      & 20  & 87 & 349 & -1.50382815 & -1.37827038 \\
RaF \\
G & TZ & dyall.cv3z & cc-pCVTZ & -30  & 500 & 61 & 415 & -2.09671991 & -1.91684123 \\
H & TZ & dyall.cv3z & cc-pCVTZ & -130 & 500 & 79 & 415 & -2.77624243 & -2.55153111 \\
I & TZ & dyall.cv3z & cc-pCVTZ &      & 500 & 97 & 415 & -3.13733209 & -2.89923481 \\
J & QZ & dyall.cv4z & cc-pCVQZ & -30  & 20  & 61 & 449 & -1.76368821 & -1.63988444 \\
K & QZ & dyall.cv4z & cc-pCVQZ & -130 & 20  & 79 & 449 & -1.82908547 & -1.69728677 \\
L & QZ & dyall.cv4z & cc-pCVQZ &      & 20  & 97 & 449 & -1.83544557 & -1.70314714\\
\end{tabular}
\end{center}
\end{ruledtabular}
\label{basis}
\end{table*}
\par
The matrix element of scalar-pseudoscalar P,T-odd interaction constant, $W_{\text{s}}$, is given by
\begin{eqnarray}
 W_{\text{s}}=\frac{1}{\Omega k_\text{s}}\langle \Psi_{\Omega}|\sum_j^n H_{\text{SP}}(j)| \Psi_{\Omega} \rangle,
\label{W_s}
\end{eqnarray}
where, $k_s$ is the dimension less electron-nucleus scalar-pseudoscalar coupling constant which is 
defined as Z$k_s$=(Z$k_{s,p}$+N$k_{s,n}$), where $k_{s,p}$ and $k_{s,n}$
are electron-proton and electron-neutron coupling constant, respectively.\par
The interaction Hamiltonian is defined as \cite{hunter_1991}
\begin{eqnarray}
H_{\text{SP}}= i\frac{G_{F}}{\sqrt{2}}Zk_{s} \gamma^0 \gamma^5 \rho_N(r) ,
\label{H_SP}
\end{eqnarray}
where $\rho_N(r)$ is the nuclear charge density normalized to unity
and G$_F$ is the Fermi constant.
The calculation of the above matrix elements depends on the accurate wavefunction in the core (near nuclear) region and the
standard way to determine the accuracy of the electronic wavefunction in that region is to compare the theoretically
calculated hyperfine structure (HFS) constant with the experimental value.
The magnetic hyperfine constant of the $J^{th}$ electronic state of an atom is given by
\begin{eqnarray}
 A_J = \frac{\vec{\mu_k}}{IJ} \cdot \langle \Psi_J | \sum_i^n \left( 
       \frac{\vec{\alpha}_i \times \vec{r}_i}{r_i^3} \right) | \Psi_J \rangle,
 \label{hfs_atom}
\end{eqnarray}
where $\Psi_J$ is the wavefinction of the $J^{th}$ electronic state, $I$ is the nuclear spin quantum number
and $\vec{\mu}_k$ is the magnetic moment of the nucleus $k$.
For a diatomic molecule, The parallel ($A_{\|}$) and perpendicular ($A_{\perp}$) magnetic hyperfine constant
of a diatomic molecule can be written as
\begin{eqnarray}
A_{\|(\perp)}= \frac{\vec{\mu_k}}{I\Omega} \cdot \langle \Psi_{\Omega} | \sum_i^n
\left( \frac{\vec{\alpha}_i \times \vec{r}_i}{r_i^3} \right)_{z(x/y)} | \Psi_{\Omega(-\Omega)}  \rangle,
\label{hfs_mol}
\end{eqnarray}
where the value of $\Omega$ is 1/2 for the ground electronic state ($^{2}\Sigma$) of RaF. \par
\section{Computational details}\label{comp}
The locally modified version of DIRAC10 \cite{dirac10} program package is used to solve the DHF equation and to construct the one-body, two-body
matrix elements and the one electron property integrals of interest. Finite size of nucleus with Gaussian charge distribution
is considered as the nuclear model where the nuclear parameters \cite{visscher_1997} are taken as default values of DIRAC10.
Small component basis functions are generated from the large component by applying restricted kinetic balance (RKB) \cite{dyall_2007}
condition. The basis functions are represented in scalar basis and unphysical solutions are removed by means of the
diagonalization of free particle Hamiltonian. This generates the electronic and positronic solution in 1:1 manner.
In our calculations, we have used the following uncontracted basis sets: triple zeta (TZ) basis: dyall.cv3z \cite{dyall_s} for Ra
and cc-pCVTZ \cite{ccpcvxz_b-ne} for F; quadruple zeta (QZ) basis: dyall.cv4z \cite{dyall_s} basis for Ra and cc-pCVQZ \cite{ccpcvxz_b-ne} basis for F.
In TZ basis, three calculations are done for the magnetic HFS constant of Ra$^{+}$ by using 51, 69 and 87 number
of correlated electrons and these are denoted by A, B and C, respectively. In QZ basis, three more calculations are
done by using 51, 69 and 87 number of correlated electrons and these are denoted by D, E and F, respectively.
The properties of RaF are calculated using two different basis. In TZ basis, three calculations are done by using
61, 79 and 97 correlated electrons and those are denoted by G, H and I, respectively and similarly in QZ basis, the calculations
using 61, 79 and 97 correlated electrons are denoted by J, K and L, respectively.
The bond length of RaF is taken as 4.23$a_0$ (2.24 \AA) \cite{kudashov_2014} in all our calculation. \par
\section{Results and discussion}\label{res_dis}
\begin{table}[b]
\caption{Hyperfine coupling constant (in MHz) of $^{223}$Ra$^{+}$}
\begin{ruledtabular}
\begin{center}
\begin{tabular}{lccr}
Basis & Expectation & Z-vector & Expt. \cite{wendt_1987, neu_1989}\\
\hline
A  & 3458 & 3418 & × \\
B  & 3504 & 3464 & × \\
C  & 3547 & 3506 & 3404(2)\\
D  & 3434 & 3394 & × \\
E  & 3448 & 3409 & × \\
F  & 3453 & 3414 & × \\
\end{tabular}
\end{center}
\label{ra_hfs}
\end{ruledtabular}
\end{table}
The aim of the present study is to exploit RaF molecule for the eEDM experiment and to provide more accurate
value of the P,T-odd interaction constants of RaF. Since, there are no experimental analogue of the
P,T-odd interaction constants like $E_\text{eff}$ and $W_\text{s}$, the accuracy of these theoretically
obtained quantities can be assessed by comparing the theoretically obtained HFS values with the corresponding experimental
values. Unfortunately, the experimental HFS results of Ra in RaF are not available. Therefore, we compare 
the experimental HFS value of $^{223}$Ra$^{+}$ \cite{wendt_1987, neu_1989} with the value obtained by theory using the same basis of Ra
as used for the calculation of RaF. \par
In Table \ref{basis}, we present the information regarding the employed basis-sets, cutoff used for occupied and virtual orbitals
and the the number of active spinor for the correlation calculation. We also compiled the correlation
energy obtained from second-order many-body perturbation theory (MBPT(2)) and CCSD method in the same table. \par
\begin{figure}[ht]
\centering
\begin{center}
\includegraphics[height=5cm, width=8cm]{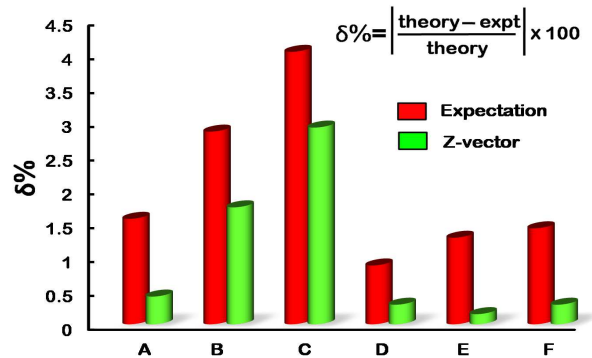}
\caption {Comparison of relative deviations between the results of expectation value and Z-vector method with experiment.}
\label{delta}
\end{center}  
\end{figure}
In Table \ref{ra_hfs}, we present the ground state ($^{2}$S) magnetic HFS constant value of $^{223}$Ra$^{+}$
using both expectation value and Z-vector method. Our results are compared with the available
experimental value \cite{wendt_1987, neu_1989}. The deviations of Z-vector and expectation values from the experiment are
presented in Fig. \ref{delta}. It is clear that the deviations of expectation value method are always
greater than those of Z-vector method. This is expected because Z-vector is a better method than the
expectation value method for the ground state property; in fact, the Z-vector value is the corresponding
expectation value plus some additional terms which make it closer to the FCI property value.
It is interesting to note that when we go from TZ to QZ basis with same number of correlated electrons
(i.e., from A to D, B to E, and C to F), the relative deviation of both Z-vector and expectation value
decreases. This is because QZ, in comparison to TZ, further improves the configuration space by adding one higher angular
momentum basis function.
It is also interesting to see that in TZ basis, if we go from A to B and B to C, the addition of 18 electrons
(4s+3d+4p and 1s-3p) changes the Z-vector HFS constant by 46 MHz and 42 MHz. Similarly in QZ basis,
as we go from D to E and E to F, the addition of 18 electrons changes the Z-vector HFS constant by
15 MHz and 5 MHz. From this observation, we can comment that the core polarization plays a definite role
in the correlation contribution of HFS constant and the effect is severe for lower basis sets.
Further, the enlargement of basis set and addition of core electrons have opposite effects in 
the calculated HFS value of Ra$^{+}$.
However, The magnetic HFS constant obtained in all electron Z-vector calculation using QZ basis (basis F)
is very close to the experimental value ($\delta$\% = 0.29).\par
\begin{table}[ht]
\caption{ Molecular dipole moment ($\mu$) and magnetic HFS constants of $^{223}$Ra in RaF }
\begin{ruledtabular}
\newcommand{\mc}[3]{\multicolumn{#1}{#2}{#3}}
\begin{center}
\begin{tabular}{lcccccr}
Basis & \mc{2}{c}{$\mu$ (D)} & \mc{2}{c}{A$_{\perp}$ (MHz)} & \mc{2}{c}{A$_{\|}$ (MHz)} \\
\cline{2-3} \cline{4-5} \cline{6-7}
× & Expect. & Z-vector & Expect. & Z-vector & Expect. & Z-vector \\
\hline
G & 3.7059 & 3.7220 & 2031 & 1987 & 2123 & 2078 \\
H & 3.7028 & 3.7207 & 2059 & 2014 & 2152 & 2107 \\
I & 3.7017 & 3.7201 & 2084 & 2038 & 2178 & 2132 \\
J & 3.8404 & 3.8474 & 2029 & 1982 & 2119 & 2072 \\
K & 3.8375 & 3.8459 & 2037 & 1991 & 2128 & 2082 \\
L & 3.8374 & 3.8459 & 2040 & 1993 & 2131 & 2085 \\
\end{tabular}
\end{center}
\end{ruledtabular}
\label{raf_hfs}
\end{table}
The properties described by Eqs. \ref{E_eff}, \ref{W_s} and \ref{hfs_mol} strongly depend on the electronic
configuration of the given (heavy) atom and are also known as ``atom in compound (AIC)'' properties \cite{AIC}.
The accuracy of the theoretically calculated AIC properties depends on the accurate evaluation of the electron density
near the atomic core region. From the accuracy of our calculated HFS constant of Ra$^{+}$ ($\delta$\% = 0.29), we can comment
that the all electron Z-vector calculation produces an accurate wavefunction in the vicinity of Ra nucleus
and we also expect the same kind of accuracy for RaF molecule. \par
We have calculated the molecular-frame dipole moment ($\mu$) of RaF, perpendicular (A$_{\perp}$)
and parallel (A$_{\|}$) magnetic HFS constants of
$^{223}$Ra in RaF using both expectation value and Z-vector method. The results are compiled in 
Table \ref{raf_hfs}. From this table, it is clear that inclusion of more core electrons decreases the
value of $\mu$ but increases the value of magnetic HFS constants of $^{223}$Ra in RaF. On the other hand,
if we go from TZ to QZ basis, the $\mu$ value is increased but the magnetic HFS values are decreased.
This observation shows that the increase of correlation space either by the addition of core
electrons or higher angular momentum wavefunctions have opposite effect on the near nuclear and outer region
part of the molecular wavefunction of RaF. We can also comment that the enlargement of basis set and core
electrons have opposite effects in the properties of RaF.
\begin{table}[ht]
\caption{P,T-odd interaction constants and their ratio of RaF}
\begin{ruledtabular}
\newcommand{\mc}[3]{\multicolumn{#1}{#2}{#3}}
\begin{center}
\begin{tabular}{lcccccr}
Basis  & \mc{2}{c}{W$_\mathrm{s}$ (kHz)} & \mc{2}{c}{E$_\mathrm{eff}$ (GV/cm)} & \mc{2}{c}{R (10$^{18}$/e cm)}\\
\cline{2-3} \cline{4-5} \cline{6-7}
× & Expect. & Z-vector & Expect. & Z-vector & Expect. & Z-vector\\
\hline
G & 144.7 & 143.6 & 53.9 & 53.5 & 90.1 & 90.1 \\
H & 147.4 & 146.3 & 54.9 & 54.5 & 90.1 & 90.1 \\
I & 149.3 & 148.1 & 55.6 & 55.1 & 90.0 & 90.0 \\
J & 141.2 & 140.4 & 52.6 & 52.3 & 90.1 & 90.1 \\
K & 141.9 & 141.1 & 52.8 & 52.5 & 90.0 & 90.0 \\
L & 142.0 & 141.2 & 52.8 & 52.5 & 89.9 & 89.9
\end{tabular}
\end{center}
\end{ruledtabular}
\label{raf_pt}
\end{table}
\par
In Table \ref{raf_pt}, we present the two P,T-odd interaction constant, namely E$_\mathrm{eff}$ and
W$_\mathrm{s}$. The E$_\mathrm{eff}$ value of RaF in QZ basis using all electron Z-vector calculation
(basis L) is 52.5 GV/cm. This E$_\mathrm{eff}$ value of RaF is even higher than the E$_\mathrm{eff}$ value of
YbF in its ground state \cite{koxlov_1994, kozlov_1997, titov_1996, quiney_1998, parpia_1998, mosyagin_1998}.
The W$_\mathrm{s}$ value of RaF using Z-vector method in the same basis (QZ, all electron) is 141.2 kHz.
This high value of W$_\mathrm{s}$ suggests that the S-PS interaction will also be responsible for significant
change in the P,T-odd frequency shift in the eEDM experiment.
These results reveal the possibility of using RaF in future eEDM experiment.
The ratio (R) of E$_\mathrm{eff}$ to W$_\mathrm{s}$ is also calculated as this is a very important quantity to obtain
the independent limit of d$_e$ and k$_s$ by using two independent experiments. Our calculated value of R
using all electron Z-vector method in QZ (L) basis is 89.9 in units of 10$^{18}$/e cm. Using this ratio, the relation of
independent d$_e$ and k$_s$ with experimentally determined d$_e^{expt}$ becomes (for more details see Ref. \cite{sasmal_hgh})
\begin{eqnarray}
 d_e + 5.56 \times 10^{-21} k_s = d_e^{expt}|_{\!_{k_s=0}},
\label{relation}
\end{eqnarray}
where $d_e^{expt}|_{\!_{k_s=0}}$ is the eEDM limit derived from the experimentally measured P,T-odd frequency shift at the limit k$_s$ = 0.
\begin{table}[ht]
\caption{ Comparison of magnetic HFS constant ($^{223}$Ra), W$_\mathrm{s}$ and E$_\mathrm{eff}$ of RaF }
\begin{ruledtabular}
\newcommand{\mc}[3]{\multicolumn{#1}{#2}{#3}}
\begin{center}
\begin{tabular}{lcccr}
Method & A$_{\perp}$ & A$_{\|}$ & W$_\mathrm{s}$ & E$_\mathrm{eff}$ \\
× & (MHz) & (MHz) & kHz & (GV/cm) \\
\hline
ZORA-GHF \cite{isaev_2013}  & 1860 & 1900 & 150 & 45.5 \\
SODCI \cite{kudashov_2014}  & 1720 & 1790 & 131 & 49.6 \\
FS-RCC \cite{kudashov_2014} & 2020 & 2110 & 139 & 52.9 \\
This work (QZ basis, & all electron) & \\
Expect. & 2040 & 2131 & 142.0 & 52.8 \\
Z-vector & 1993 & 2085 & 141.2 & 52.5
\end{tabular}
\end{center}
\end{ruledtabular}
\label{comparison}
\end{table}
\par
We have compared our calculated results with other theoretically obtained values in table \ref{comparison}. The first {\it ab initio} calculation
of W$_\mathrm{s}$ of RaF was performed by Isaev {\it et al.} \cite{isaev_2013}. They employed two-component zeroth-order regular
approximation (ZORA) generalized Hartree-Fock (GHF) method and obtained the value of W$_\mathrm{s}$
as 150 kHz. They also obtained the value of E$_\mathrm{eff}$ as 45.5 GV/cm by using ZORA-GHF value of
W$_\mathrm{s}$ and the approximate ratio between E$_\mathrm{eff}$ and W$_\mathrm{s}$.
Kudashov {\it et al.} \cite{kudashov_2014} employed two different methods to incorporate relativistic and electron correlation
effects: (i) spin-orbit direct configuration interaction (SODCI) method and (ii) relativistic two-component
Fock-space coupled cluster approach (FS-RCC) within single- and double- excitation approximation.
However, it is worth to remember that truncated CI is not size extensive and thus cannot treat
electron correlation properly, specially, for the heavy electronic system like RaF where the number of electron
is so large. In their FS-RCC method, Kudashov {\it et al.} \cite{kudashov_2014} calculated the properties of RaF using the
finite field method, which is a numerical technique. They corrected the error associated with their
calculation considering higher order correlation effect and basis set with the addition of partial triple in the CCSD model (CCSD(T))
and using enlarged basis set, respectively. They have done those corrections only for the ground state
((0,0) sector of Fock space) coupled cluster amplitudes.
On the other hand, we have calculated the property values of RaF via two analytical methods (expectation value
and Z-vector method) in the relativistic coupled-cluster framework within four-component formalism.
We also calculated the E$_\text{eff}$ and W$_\text{s}$ values directly by using 
Eqs. \ref{E_eff} and \ref{W_s}, respectively. \par
\section{Conclusion}\label{conc}
In conclusion, we have applied both Z-vector and expectation value method in the relativistic coupled-cluster framework
to calculate parallel and
perpendicular magnetic HFS constant of $^{223}$Ra in RaF, E$_\mathrm{eff}$ and W$_\mathrm{s}$ of RaF.
We have also calculated the magnetic HFS constant of $^{223}$Ra$^{+}$ to show the reliability of our
results. Our most reliable value of E$_\mathrm{eff}$ and W$_\mathrm{s}$ of RaF are 52.5 GV/cm and
141.2 kHz, respectively. This shows that RaF can be a potential candidate for eEDM experiment.
We also showed that core electrons play significant role and the effect is notable for lower basis sets.
Our results also show that the Z-vector, being an energy derivative method, is much more reliable than
the expectation value method.
\section*{Acknowledgement}
Authors acknowledge a grant from CSIR 12th Five Year Plan project on Multi-Scale Simulations of Material (MSM)
and the resources of the Center of Excellence in Scientific Computing at CSIR-NCL. S.S. and H.P acknowledge the CSIR
for their fellowship.
S.P. acknowledges funding from J. C. Bose Fellowship grant of Department of Science and Technology (India).

\end{document}